%% file: Project.tex
\documentclass[review,final]{elsarticle}

\journal{International Journal of Heat and Mass Transfer}

\usepackage[utf8]{inputenc}
\usepackage[russian,english]{babel}
\usepackage{bm}
\usepackage{amssymb}
\usepackage{amsxtra}
\usepackage{amsthm}
\usepackage[mathscr]{eucal}

\usepackage{xcolor}
\usepackage{graphicx}

\usepackage{mathtools}
\usepackage{bigints}

\usepackage[colorlinks,allcolors=blue,unicode]{hyperref}

\hypersetup{
    pdftitle = {Heat conduction in 1D harmonic crystal},
}

\input def

\begin{document}

\input{comment}

\end{document}

%% file: def.tex
\def\I{\mathrm{i}}

\def\be#1{\begin{equation}\label{#1}}
\def\ee{\end{equation}}
\newcommand {\ba}[2]{\be{#1}\begin{array}{#2}}
\newcommand {\ea}{\end{array} \ee}

\renewcommand{\=}{\stackrel{\mbox{\scriptsize def}}{=}}

\def\({\left(}
\def\){\right)}

\def\d{{\cal D}}

\def\sign{\operatorname{sign}}

\def\d{\mathrm d}

\def\tau{{t}}

\theoremstyle{remark}
\newtheorem{remark}{Remark}

%% file: comment.tex

\begin{frontmatter}

\selectlanguage{english}
\title{Discrete and continuum fundamental solutions describing heat conduction
in a 1D harmonic crystal: Discrete-to-continuum
limit and slow-and-fast motions decoupling}
\author{Serge N.~Gavrilov}
\ead{serge@pdmi.ras.ru}
\address{Institute for Problems in Mechanical Engineering RAS, V.O., Bolshoy
pr. 61, St.~Petersburg, 199178, Russia}

\begin{abstract}        
%
In the recent paper by Sokolov et al.\ (Int. J. of Heat and Mass Transfer 176,
2021, 121442)
ballistic heat propagation in a 1D harmonic crystal is considered
and 
the properties of the exact discrete solution and the continuum solution of the ballistic
heat equation 
are numerically compared.
The aim of this note is to demonstrate that the continuum fundamental solution 
can be formally obtained as the slow time-varying component of the 
large-time asymptotics for the exact discrete solution on a
moving point of observation.
\end{abstract}

\begin{keyword}	
ballistic heat transfer
\sep
harmonic crystal
\sep
asymptotics
\sep
the method of stationary phase
\end{keyword}

\end{frontmatter}

\section{Introduction}

In recent paper~\cite{Sokolov} 
ballistic heat propagation in a 1D harmonic crystal is considered, 
and the properties of the exact discrete solution for the kinetic temperature 
and the approximate continuum solution
are numerically compared. The discrete solution and the continuum one can be obtained as 
the convolutions (see
Sect.~\ref{intro}) of the initial conditions with the corresponding
fundamental solutions.
The exact discrete fundamental solution is
\cite{schrodinger1914dynamik,klein1953mecanique,hemmer1959dynamic}
\begin{equation}
\tilde T_n(\tau)\=2J_{2n}^2(2\tau).
\label{sol-discrete}
\end{equation}
The continuum fundamental solution
\cite{krivtsov2015heat,krivtsov-da70,gavrilov2018heat}
\begin{equation}
\tilde T({x},\tau)\=
\frac{H(\tau-|{x}|)}{\pi\sqrt{\tau^2-{x}^2}}
\label{sol-continuum}
\end{equation}
satisfies a partial differential equation 
called the ballistic heat equation
\cite{krivtsov-da70}.
Here $J_{2n}(\cdot)$ is the Bessel function 
of the
first kind of integer order $2n$, $n$ is a particle number, $H(\cdot)$ is the Heaviside function, $t$ is the dimensionless time, $x$ is the
dimensionless spatial co-ordinate ($x=n$ for any integer $x$).
The ballistic heat equation was introduced by
Krivstov 
\cite{krivtsov2015heat}, who considered the infinite system of 
differential-difference equations for covariance variables and applied the
procedure of continualization.

The aim of this note is to demonstrate that the continuum fundamental solution 
\eqref{sol-continuum} can be formally obtained as the slow time-varying component of the 
large-time asymptotics for the exact discrete solution \eqref{sol-discrete} on a
moving point of observation.

\section{Mathematical formulation}
\label{intro}
In this section we briefly formulate the problem concerning an initial random
excitation for a 1D harmonic crystal in the framework of the two approaches to
introduce the fundamental solutions, which we plan to compare.

\subsection{The discrete (exact) approach}
Both solutions
\eqref{sol-discrete}
and
\eqref{sol-continuum}
describe the propagation of the kinetic temperature in the same
infinite mechanical system, governed by the following equations and initial
conditions:
\begin{gather}  
\ddot u_n= u_{n+1}-2u_n+u_{n-1},
\label{sol-maineq}
\\
u_n(0)=0,\qquad \dot u_n(0)=\rho_n.
\label{sol-ic}
\end{gather}
Here $n\in\mathbb Z$, $\rho_n$ are uncorrelated random quantities such that 
\begin{gather}
\langle\rho_n\rangle=0,\qquad \langle\rho_n\rho_k\rangle=\sigma_n\delta_{nk};
\label{sol-cov-general}
\end{gather}
overdot denotes the derivative with respect to dimensionless time $\tau$, $\delta_{nk}$ is the Kronecker delta, the angle brackets denote the
mathematical expectation.
The kinetic temperature is conventionally introduced by the following formula
\begin{equation}
T_n\=
2k_B^{-1}\langle K_n\rangle,
\label{sol-T-def}
\end{equation}
where 
$
K_n(\tau)=\frac{\dot u_n^2(\tau)}2
$
is the kinetic energy of the particle with number $n$, $k_B$ is
the (dimensionless) Boltzmann constant.

\begin{remark}  
In this paper we discuss mostly the fundamental solutions and consider 
a slightly different and simpler problem
formulation than the one used 
in \cite{Sokolov}. In \cite{Sokolov}, following to \cite{hemmer1959dynamic},
both the kinetic and the potential energy were initially equally excited, and,
therefore, both initial conditions 
\eqref{sol-ic} were non-zero.
\end{remark}

Consider the particular case for $\rho_n$, namely, a random point excitation: 
\begin{equation}
\rho_n=\rho_0\delta_{n0}.
\label{sol-ranpoi}
\end{equation}
The exact expression for the particle velocity is 
\cite{schrodinger1914dynamik,klein1953mecanique,hemmer1959dynamic}
\begin{equation}
\dot u_n=\rho_0 J_{2n}(2\tau).
\end{equation}
Accordingly, the exact expression for the mathematical expectation of the kinetic 
energy is
\begin{equation}
\langle K_n(\tau)\rangle=
\frac12\,\langle \dot u^2_n \rangle 
=
\mathscr E_0\, J^2_{2n}(2\tau),
\label{sol-K-exact}
\end{equation}
where 
\begin{gather}  
\mathscr E_0\=\sum_{n=-\infty}^{\infty} \langle K_n(0)\rangle=\frac{\sigma_0}2
\label{sol-eps0-def}
\end{gather}
is the mathematical expectation for the initial kinetic (as well as the total)
energy for the whole crystal in the case of point excitation. Thus, since 
\begin{equation}
J^2_{2n}(0)=\delta_{n0},       
\end{equation}
formulas 
\eqref{sol-T-def} and  
\eqref{sol-K-exact}
result in 
\begin{gather}  
T_n(\tau)
={k_B^{-1}\mathscr E_0}\tilde T_n(\tau),
\label{sol-dis-pre}
\end{gather}
where $\tilde T_n(\tau)$ (the discrete fundamental solution)  
is defined by Eq.~\eqref{sol-discrete}. 
For $\tau=0$ the last formula reads 
\begin{gather}  
T_n(0)
={2k_B^{-1}\mathscr E_0} \delta_{n0}=k_B^{-1}\sigma_0 \delta_{n0}.
\end{gather}

In more general case \eqref{sol-cov-general}, the solution of problem 
\eqref{sol-maineq}--\eqref{sol-cov-general} can be expressed in the form of 
the discrete spatial convolution:
\begin{equation}
T_n(\tau)=\frac{k_B^{-1}}2\sigma_n\star\tilde T_{n}(\tau)\=
\frac{k_B^{-1}}2\sum_{k=-\infty}^\infty \sigma_k\tilde T_{n-k}(\tau).
\label{sol-exact-convolution}
\end{equation}
It is known 
\cite{klein1953mecanique,krivtsov2014energy,Gavrilov2019}
that in the case $\sigma_n=\mathrm{const}$ exact solution \eqref{sol-exact-convolution}
of problem \eqref{sol-maineq}--\eqref{sol-cov-general} describes 
the process of thermal equilibration of the kinetic energy $K\equiv K_n$ and
the potential
one 
\begin{equation}
\Pi\equiv\Pi_n=\frac12\big\langle
(u_{n+1}-u_n)^2
\big\rangle
.
\end{equation}
Namely, in the last case, according to 
\eqref{sol-exact-convolution} {one has}
\cite{klein1953mecanique,krivtsov2014energy,Gavrilov2019}
\begin{equation}
\mathscr L=K-\Pi=
O(\tau^{-1/2}),
\label{sol-equilibration}
\end{equation}
where $\mathscr L$ is the Lagrangian.

\subsection{The continuum (approximate) approach}
The kinetic temperature propagation in the system described by 
Eqs.~\eqref{sol-maineq}--\eqref{sol-ic}
can be {approximately} described by the ballistic heat equation
\cite{krivtsov2015heat,krivtsov-da70}:
\begin{equation}
\ddot T(x,\tau)+\frac1\tau\,\dot T(x,\tau)=T''(x,\tau).
\label{sol-thermocon}
\end{equation}
Here $T({x},\tau)$ is the kinetic temperature per unit length (a continuum
quantity), prime denotes the spatial derivative with respect to ${x}$.
The corresponding initial conditions are
\begin{equation}
T({x},0)=T_0({x}),\qquad \dot T({x},0)=0.
\label{sol-thermocon-ic}
\end{equation}
The initial temperature $T_0({x})$ is assumed to be a slowly varying function.
The solution of Eqs.~\eqref{sol-thermocon}--\eqref{sol-thermocon-ic}
can be expressed in the form of a spatial convolution 
\cite{krivtsov2015heat,krivtsov-da70}:
\begin{equation}
T({x},\tau)=T_0({x})\ast\tilde T({x},\tau) \=\int_{-\infty}^{+\infty}
T_0(\xi)\tilde T({x}-\xi,\tau)\,\d\xi,
\label{sol-approximate-convoluton}
\end{equation}
where 
$\tilde T$ is the continuum fundamental solution.
Consider the case of a point excitation, i.e.,
the solution of the ballistic heat equation
with initial conditions 
\begin{gather}
T({x},0)=T_0^0\delta({x}),\qquad \dot T({x},0)=0.
\label{sol-thermocon-ic-delta}
\end{gather}
The mathematical expectation for the initial kinetic energy of the whole crystal in the
framework of the continuum approach is
\begin{equation}
\frac{k_B}2\int_{-\infty}^{\infty} T_0({x})\,\d {x}
=\frac{k_BT_0^0}2.
\end{equation}
The value of $T_0^0$ should be chosen in order to make problem 
\eqref{sol-thermocon}--\eqref{sol-thermocon-ic}
physically equivalent to 
\eqref{sol-maineq},
\eqref{sol-ic},
\eqref{sol-ranpoi}.
This requirement is essential to get the approximate continuum solution 
\eqref{sol-approximate-convoluton}
close to the exact solution~\eqref{sol-exact-convolution}. 
The continuum approach implicitly assumes that
\begin{itemize}
\item process of thermal equilibration in the case of slowly varying $T_0({x})$
is close to one observed in the case of constant $T_0({x})$ (see 
\eqref{sol-equilibration});
\item
the ballistic heat equation \eqref{sol-thermocon} becomes valid only for large
times after
equilibration, when $\mathscr L\simeq0$.
\end{itemize}
Accordingly, for the same physical problem 
{\it the initial kinetic energy for the whole crystal, calculated
in the framework of the continuum approach 
should be equal to a half of the initial kinetic energy of the whole crystal
observed in the 
framework of the exact discrete approach.}
In particular, considering the case of a random point excitation
\eqref{sol-ranpoi} in the framework of the continuum approach,
we need to take 
initial conditions in the form of 
\eqref{sol-thermocon-ic-delta}, where 
$T_0^0$ is such that
$
\frac{k_BT_0^0}2 
=\frac{\mathscr E_0}2.
$
Thus, 
the continuum solution, which corresponds to the discrete solution 
\eqref{sol-dis-pre}, is
\begin{gather}
T({x},t)=
T_0^0
\tilde T({x},t)
=
{k_B^{-1}\mathscr E_0}
\tilde T({x},\tau)
.
\label{sol-cont-sol-final}
\end{gather}

Looking at Eqs.~\eqref{sol-dis-pre} and 
\eqref{sol-cont-sol-final} one can see that 
to compare the solutions obtained in the frameworks of the discrete and
continuum approaches one needs to compare the fundamental solutions 
\eqref{sol-discrete} and \eqref{sol-continuum}.
Note that since the initial temperature in the form of the first equation 
in \eqref{sol-thermocon-ic-delta} is not a slowly varying function, the
continuum approach is not applicable for a point source, and the
solutions, generally speaking, are not close to each other. Indeed, in the
case of point excitation, the local energy equilibration does not take place.
In such a case we can speak about 
the energy equilibration for the whole crystal only. 

\section{Asymptotics}
Now we want to show that  
$\tilde T(x,\tau)$ 
can be formally obtained as a slow component of
large-time asymptotics of the corresponding exact discrete solution on a
moving point of observation, and
looks in some sense like a spatial average of $\tilde T_n(\tau)$.
The discrete solution
\eqref{sol-discrete}
has a physical meaning only for $n\in\mathbb Z$. Thus, we can express 
\eqref{sol-discrete} in terms of the Anger function $\mathbf J_n(\tau)$%
\footnote{The question what is the best continuum approximation for
a solution defined only at integer values of a spatial co-ordinate is discussed in book by Kunin
\cite{Kunin1982}}
\cite{abramowitz1972handbook}:
\begin{gather}
\mathbf J_n(\tau)=J_n(\tau),\quad n\in\mathbb Z,
\label{sol-JJ}\\
\mathbf J_n(\tau)\=\frac1{2\pi}\int_{-\pi}^{\pi}\exp \I(\tau\sin\omega-n\omega)\,\d\omega,
\quad n\in\mathbb R.
\label{sol-J}
\end{gather}
Formula \eqref{sol-discrete} defines an even function of $n$. Taking into
account
Eq.~\eqref{sol-JJ},
we can rewrite formula \eqref{sol-discrete} as follows:
\begin{equation}
\tilde T_n(\tau)=2\mathbf J_{2|n|}^2(2\tau).
\label{sol-discrete1}
\end{equation}

\begin{figure}[p]	
\centering\includegraphics[width=0.9\columnwidth]{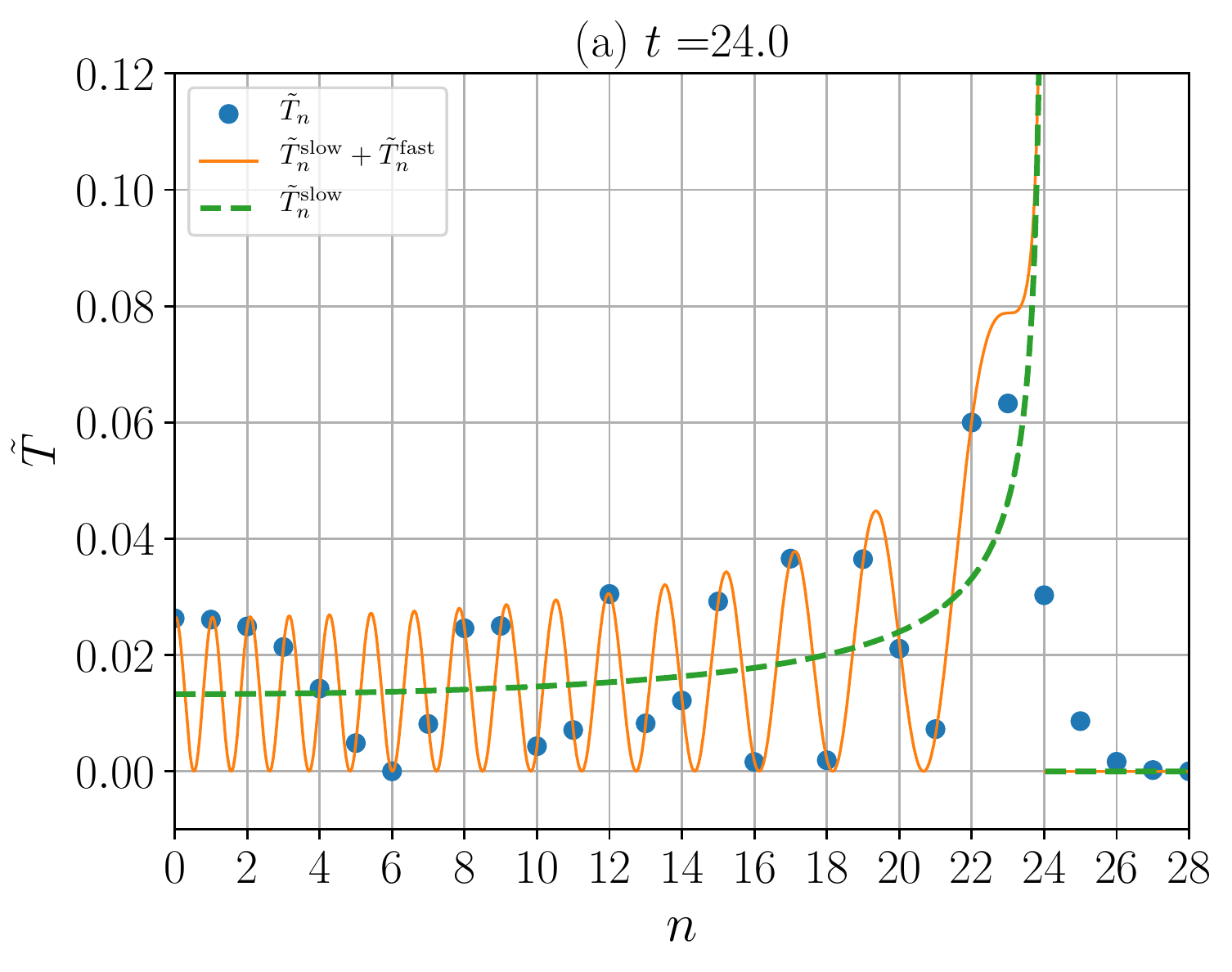}
\centering\includegraphics[width=0.9\columnwidth]{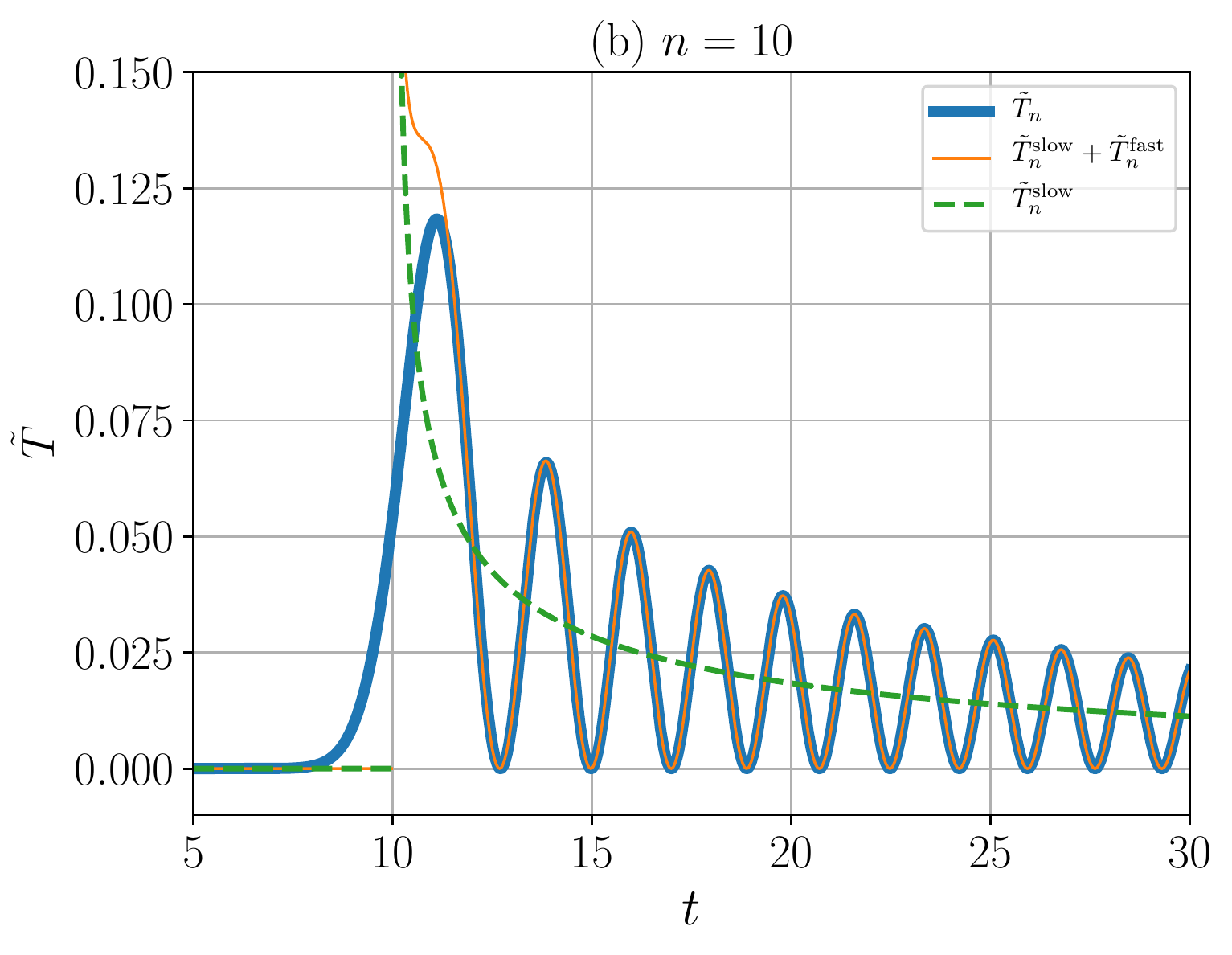}
\caption{Comparing the $\tilde T_n(\tau),\  
\tilde T_n^{\mathrm{slow}}(\tau)+\tilde T_n^{\mathrm{fast}}(\tau)$ and 
$\tilde T_n^{\mathrm{slow}}(\tau)$. (a) The kinetic temperature versus the spatial
co-ordinate $n$, (b) the kinetic temperature versus the time $t$.}
\label{com-fig-comp.eps}
\end{figure}

Let us calculate the large-time asymptotics 
of the right-hand side of \eqref{sol-discrete1} on the moving
front\footnote{This approach \cite{Slepian1972} allows one to
describe running waves, wave-fronts, and describe the wave-field as a whole.}
\begin{equation}
|n|=V\tau,\quad V=\mathrm{const},\quad \tau\to\infty
,\quad\tau\in\mathbb R
,\quad n\in\mathbb R
\label{sol-front1}
\end{equation}
considering $n$ as a continuum spatial variable.
Here the meaning of the quantity $V\geq0$ is the velocity for the observation point. 
To estimate the right-hand side of 
\eqref{sol-J} 
we now use the method of stationary phase
\cite{Fedoruk-Saddle}. One has
\begin{gather}
\mathbf J_{V\tau}(\tau)=\frac1{2\pi}\int_{-\pi}^{\pi}
\exp \I \tau\phi(\omega)
\,\d\omega,
\quad \tau\to\infty; 
\label{Anger-i} 
\\ 
\phi(\omega)\=\sin\omega-V\omega.  
\end{gather} The stationary points for the phase function $\phi(\omega)$ 
are defined by the condition $\phi'=0$. There are no stationary points 
in the case $V>1$, therefore integral 
\eqref{Anger-i} can be roughly estimated as $O(t^{-1})$.
In the case $0\leq V<1$ the stationary points are solutions of the equation
$
\cos \omega=V,
$
or, in the explicit form, 
\begin{equation}
\omega_\pm=\pm\arccos V.
\end{equation}
One gets
\begin{gather}
\phi(\omega_\pm)=\sin\omega_\pm-\omega_\pm=\pm
{\sqrt{1-V^2}}
-V\arccos V
,
\\
\phi''=-
\sin\omega,\\
\phi''(\omega_\pm)=\mp
{\sqrt{1-V^2}}.
\end{gather}
Now using the formula for contribution from a stationary point
\cite{Fedoruk-Saddle},  in the case $0<V<1$ we obtain:
\begin{multline}
\mathbf J_{V\tau}(\tau)
=\frac1{2\pi}\sum_{(\pm)}\sqrt{\frac{2\pi}{\tau\phi''(\omega_\pm)}}
\,
\exp \I\left(\phi(\omega_\pm)\tau+\frac\pi4\sign\phi''(\omega_\pm)\right)
+O(\tau^{-1})
\\=
\sqrt\frac{2}{\pi \tau \sqrt{1-V^2}}
\,
\cos\left(
\big(
\sqrt{1-V^2}
-V\arccos V
\big)
\tau-\frac\pi4
\right)
+O(\tau^{-1}).
\end{multline}
Thus, according to \eqref{sol-discrete1}, and provided that 
\eqref{sol-front1} is true, 
one has
\begin{multline}
\tilde T_n(\tau)=
\frac{2}{\pi \tau\sqrt{1-V^2}}\,
\cos^2\left(
\left(
{\sqrt{1-V^2}}
-V\arccos V
\right)2\tau-\frac\pi4
\right)
+O(\tau^{-3/2})
\\=
\frac{1}{\pi \tau\sqrt{1-V^2}}
\left(1+
\sin
\big(
\big(
{\sqrt{1-V^2}}
-V\arccos V
\big)4\tau\big)\right)
+O(\tau^{-3/2}).
\end{multline}
Now we return to variables $n$, $t$, and substitute $V=|n|/\tau$ into the last expression. This yields 
\begin{gather}
\tilde T_n(\tau)\simeq
\tilde T_n^{\mathrm{slow}}(\tau)+\tilde T_n^{\mathrm{fast}}(\tau)
,
\quad |n|<\tau;
\label{com-decoupling}
\\
\tilde T_n^{\mathrm{slow}}(\tau)=\frac{1}{\pi\sqrt{\tau^2-n^2}},
\label{com-slow}
\\
\tilde T_n^{\mathrm{fast}}(\tau)=\tilde T_n^{\mathrm{slow}}(\tau)
\sin
\left(
\left(
\frac{\sqrt{\tau^2-n^2}}{\tau}-
\frac {|n|}\tau \arccos \frac {|n|}\tau
\right)4\tau\right)
.
\end{gather}
Formula \eqref{com-decoupling} yields the asymptotic decoupling of thermal
motions as the sum of the slow and the fast motions. The right-hand side of 
Eq.~\eqref{com-slow} coincides with 
Eq.~\eqref{sol-continuum} provided that $n={x}$. The comparison between
$\tilde T_n(\tau),\  
\tilde T_n^{\mathrm{slow}}(\tau)+\tilde T_n^{\mathrm{fast}}(\tau)$ and 
$\tilde T_n^{\mathrm{slow}}(\tau)$ is given in Fig.~\ref{com-fig-comp.eps}.

\section{Conclusion}
Up to nowadays it was unclear how to analytically derive 
the expression \eqref{sol-continuum} for the fundamental solution of the
ballistic heat equation  \eqref{sol-thermocon} basing on the fundamental
solution \eqref{sol-discrete} of the discrete problem. In the paper we have demonstrated 
that Eq.~\eqref{sol-continuum} can be formally obtained as the slow time-varying component of the 
large-time asymptotics for the exact discrete solution \eqref{sol-discrete} on a
moving point of observation.
We also
provide the direct procedure to uncouple the slow and the fast thermal motions caused
by a point heat source in a 1D harmonic crystal.

We expect that the similar approach can be applied to more complicated
and physically significant systems, e.g., to obtain the expression for the
slow motion related
to ballistic
thermal transport in a 1D harmonic crystal with an isotopic defect (such a
model has been used in \cite{Gendelman_2021} to describe the Kapitza thermal resistance).

\section*{Declaration of Competing Interests}
None to declare.

\section*{Acknowledgements} The author is grateful to A.M.~Krivtsov,
O.V.~Gendelman, 
A.A.~Sokolov, E.V.~Shishkina, A.S.~Murachev and V.A.~Kuzkin for useful and stimulating discussions.

This work is supported by Russian Science Support Foundation (Grant No. 21-11-00378).

\bibliographystyle{elsarticle-num}
\bibliography{math,thermo,serge-gost,discrete}